\documentclass[a4paper,10pt]{article}
\usepackage{amsmath}
\usepackage{amsfonts}
\usepackage{amssymb}
\usepackage{graphicx}

\begin{document}

\title{Angular momentum conservation in measurements on spin Bose-Einstein
condensates}
\author{F. Lalo\"{e} and W.J. Mullin}
\maketitle

\begin{abstract}
We discuss a thought experiment where two operators, Alice and Bob, perform
transverse spin measurements on a quantum system; this system is initially
in a double Fock spin state, which extends over a large distance in space so
that the two operators are far away from each other. Standard quantum
mechanics predicts that, when Alice makes a few measurements, a large
transverse component of the spin angular momentum may appear in Bob's
laboratory. A paradox then arises since local angular momentum conservation
seems to be violated. It has been suggested that this angular momentum may
be provided by the interaction with the measurement apparatuses. We show
that this solution of the paradox is not appropriate, so that another
explanation must be sought. The general question is the retroaction of a
quantum system onto a measurement apparatus. For instance, when the measured
system is entangled with another quantum system, can its reaction on a
measurement apparatus be completely changed? Is angular momentum conserved
only on average over several measurements, but not during one realization of
the experiment?
\end{abstract}

\tableofcontents

\begin{center}
***********
\end{center}

\bigskip

The role of conservations laws in the process of quantum measurement, and in
particular of the retroaction of the measured system on the measurement
apparatus, has been discussed from the early days of quantum mechanics.\ A
famous example is the Einstein-Bohr debate at the fifth Solvay Congress in\
Brussels, where Einstein invented a thought experiment with a moving double
slit screen and a measurement of its momentum recoil during interaction with
the test particle \cite{Jammer}.\ Wigner also analyzed the relation between
conservation laws and measurements, emphasizing that only observables
commuting with the conserved quantities could exactly be measured \cite%
{Wigner-1952}. This line of thought was then continued by Araki and Yanase
\cite{Araki-Yanase}, Osawa \cite{Osawa}, Loveridge and Bush \cite%
{Loveridge-Bush}, and others; Aharonov and Rohrlich have emphasized in their
book \cite{Aharonov-Rohrlich} the fact that all measurements are relative
(for instance a Stern-Gerlach apparatus does not measure the spin component
along an absolute direction, but a direction that is fixed by the apparatus
itself) and its impact on Wigner's argument. Leggett and Sols \cite%
{Leggett-Sols} have discussed the spontaneous appearance of a relative phase
between two large superconductors under the effect of quantum measurement.
They have pointed out that the phase of a macroscopic current may in theory
be be determined by the interaction with a small measurement apparatus:
\textquotedblleft Can it be that by placing, let us say, a miniscule compass
needle next to the system... we can force the system to 'realize' a definite
macroscopic value of the current?\textquotedblright; of course, this seems
paradoxical: how can a very small measurement apparatus make an arbitrarily
large system to completely change state?

Bose-Einstein condensates in gases provide quantum systems offering
interesting opportunities to test the laws in quantum mechanics \cite%
{Pethick-Smith}, either in thought experiments as in the tradition created
by the Einstein-Bohr debate, or in real experiments. A recent analysis \cite%
{Hidden-phase} discusses a paradoxical thought experiment with spin
condensates extending over long distances (Fig. \ref{fig-1}).\ It assumes
that two long condensates, in the ${+}$ and $-$ spin state respectively,
overlap in two remote laboratories, where Alice and Bob perform spin
measurements.\ If the populations of the two condensates are equal, the
average of the  spin angular momentum of the system vanishes.\ Nevertheless, if
Alice performs transverse spin measurements in her laboratory and finds a
polarization is some random direction $\mathbf{u}_{\varphi }$, even if she
measures a small number of spins, a transverse angular momentum parallel to $%
\mathbf{u}_{\varphi }$\ appears instantaneously in the entire system.\ This
is in particular true in Bob's laboratory, where all particles acquire a
spin polarization that is parallel to $\mathbf{u}_{\varphi }$, even if Alice
never interacted with them. Compatibility with relativity\ (no faster than
light signalling) is maintained by the completely random direction of $%
\mathbf{u}_{\varphi }$, which ensures that no signal can be transmitted in
this way at an arbitrary distance with no delay.\ Nevertheless, the
spontaneous appearance of an angular momentum in Bob's laboratory, without
any local interaction, seems to violate angular momentum conservation.\ This
is all the more true since Alice's apparatus may have interacted with a
microscopic number of spins only, while the angular momentum appearing in Bob's
laboratory is macroscopic; where does this angular momentum come from?\ We
come back to this paradox in more detail in \S\ \ref{non-local-appearance}.

Paraoanu and Healey have studied this paradox \cite{Paraoanu-Healey} and
concluded that, \textquotedblleft while this Gedankenexperiment provides a
striking illustration of several counter-intuitive features of quantum
mechanics, it does not imply a non-local violation of the conservation of
angular momentum\textquotedblright . They emphasize that, even if Bob's
system is projected by\ Alice's measurement into an eigenstate of transverse
angular momentum, \textquotedblleft we cannot attribute physical reality
prior to its measurement, even when the state of the system is an eigenstate
of that observable... There is no angular momentum before it was
measured\textquotedblright . The purpose of the present article is to
complete this discussion: considering as suggested by Paraoanu and Healey
that only measured quantities really exist, we study combined measurements
performed by Alice and Bob and the angular momentum taken by their
respective apparatuses.\ We conclude that this is not sufficient to ensure
angular momentum conservation, so that the the paradox is not lifted; its
solution should be sought in another direction.

\section{The paradox}

\label{the-paradox}We first recall the double spin condensate paradox in
more detail than in the introduction, and then how Paraoanu and Healey
propose to solve it.

\subsection{Double Fock state and transverse spin measurements}

\label{non-local-appearance}

\begin{figure}[h]
\centering
\includegraphics[trim = 40mm 66mm 25mm 20mm,clip,width=10cm]{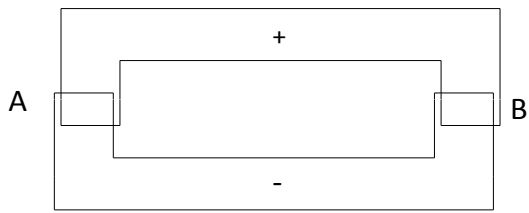}
\caption{Two highly populated Fock states associated with opposite spin
directions ($+$ and $-$) overlap in two remote regions of space, region A
where Alice operates, and region B where Bob operates.\ A series of
transverse spin measurements made by Alice triggers the appearance of a
well-defined transverse orientation in her region, and also that of a
parallel macroscopic transverse orientation in Bob's region (quantum
non-locality). Angular momentum seems to appear in region B\ from nothing,
with no interaction at all.}
\label{fig-1}
\end{figure}

Figure \ref{fig-1} schematizes the thought experiment with two spin
condensates.\ The first condensate has $N_{\alpha }$ particules occupying a
single particle state $\left\vert \alpha \right\rangle =\left\vert
u,+\right\rangle $ with orbital wave function $u$ and spin state $+$; the
second has $N_{\beta }$ particles occupying a single particle state $%
\left\vert \beta \right\rangle =\left\vert v,-\right\rangle $ with orbital
wave function $v$ and spin state $-$. The two wave functions overlap in
regions of space A and B where Alice and Bob perform measurements. The
condensates are described by Fock states (we ignore thermal effects and
condensate depletion due to interactions, assuming for instance that the
systems are dilute gases; actually, in this article, for the sake of
simplicity we take the word \textquotedblleft condensate\textquotedblright\
as equivalent to \textquotedblleft Fock state\textquotedblright). The
initial state $\left\vert \Psi \right\rangle _{0}$\ of the whole quantum
system is then:%
\begin{equation}
\left\vert \Psi \right\rangle _{0}=\left\vert N_{\alpha },N_{\beta
}\right\rangle =\frac{1}{\sqrt{N_{\alpha }!N_{\beta }!}}\left[
a_{1}^{\dagger }\right] ^{N_{\alpha }}\left[ a_{2}^{\dagger }\right]
^{N_{\beta }}\left\vert 0\right\rangle  \label{1}
\end{equation}%
Here $a_{1}^{\dagger }$ is the creation operator into the single particle
state $\left\vert u,+\right\rangle $, $a_{2}^{\dagger }$ the creation
operator into state $\left\vert v,-\right\rangle $, and $\left\vert
0\right\rangle $ the vacuum state. We call $Oz$ the quantization axis for
spins, and will consider transverse spin measurements performed by Alice and
Bob in any direction perpendicular to $Oz$.\ This state has zero average
spin angular momentum along these directions, and zero spin momentum along $%
Oz$ as well if $N_{\alpha }=N_{\beta }$.

Now Alice makes $P_{A}$ spin measurements in her laboratory, choosing
transverse directions characterized by polar angles $\varphi _{A}^{1}$, $%
\varphi _{A}^{2}$,..$\varphi _{A}^{P_{A}}$.\ These measurements are
performed with different apparatuses situated in different regions of space
of her laboratory, and they are independent; we assume that each apparatus
detects one single particle (how to include the case where no particle is
found in a particular region of space is discussed in \cite%
{Hidden-phase-Leiden}).\ Similarly, Bob performs $P_{B}$ transverse spin
measurements in his laboratory with polar angles $\varphi _{B}^{1}$, $%
\varphi _{B}^{2}$,..$\varphi _{B}^{P_{B}}$.\ As in \cite{Hidden-phase}, we
assume that the total number of measurements $P=P_{A}+P_{B}$ is much smaller
that the total number of particles $N=N_{A}+N_{B}$.\ Under these conditions,
the effects of the \textquotedblleft quantum angle\textquotedblright\ \cite%
{Mullin-Laloe}\ can be ignored and the relative phase behaves classically.\
The probability of an event where Alice obtains $m_{A}^{+}$ results $+$\ and
$m_{A}^{-}=P_{A}-m_{A}^{+}$ results $-$, while Bob obtains $m_{B}^{+}$
results $+$\ and $m_{B}^{-}=P_{B}-m_{B}^{+}$ results $-$, is then given by:%
\begin{equation}
\begin{array}{l}
\displaystyle\mathcal{P}(m_{A}^{+},m_{A}^{-},m_{B}^{+},m_{B}^{-}|\varphi
_{A}^{1},..,\varphi _{A}^{P_{A}};\varphi _{B}^{k},..,\varphi _{B}^{P_{B}})=
\\
\multicolumn{1}{r}{\displaystyle=\int_{0}^{2\pi }\frac{d\lambda }{2\pi }%
\prod_{j=1}^{m_{A}^{+}}\cos ^{2}\left( \frac{\lambda -\varphi _{A}^{j}}{2}%
\right) \prod_{j=m_{A}^{+}+1}^{m_{A}^{+}+m_{A}^{-}}\sin ^{2}\left( \frac{%
\lambda -\varphi _{A}^{k}}{2}\right) \prod_{k=1}^{m_{B}^{+}}\cos ^{2}\left(
\frac{\lambda -\varphi _{B}^{k}}{2}\right)
\prod_{k=m_{B}^{+}+1}^{m_{B}^{+}+m_{B}^{-}}\sin ^{2}\left( \frac{\lambda
-\varphi _{B}^{k}}{2}\right)}%
\end{array}
\label{2}
\end{equation}%
The interpretation of this formula is straightforward: if the two
condensates had a relative phase $\lambda $, the probability of finding a
result $+$ with an apparatus oriented along direction $\varphi $ is $\cos
^{2}\left( \lambda -\varphi \right) /2$, and that of finding the opposite
result is $\sin ^{2}\left( \lambda -\varphi \right) /2$.\ The probability of
finding the combined results is the product of the corresponding
probabilities; in this formula, Alice's results appear in the first two
factors of the product, Bob's results in the last two factors.\ But $\lambda
$ is actually completely random between $0$ and $2\pi $, so that an integral
over $\lambda $ is necessary; although the integrand is a product, this
integral introduces correlations between the successive results. This
formula is general; for instance, it remains valid if Alice only makes
measurements, not Bob; it is sufficient to put $m_{B}^{+}=m_{B}^{-}=0$, so
that the last two products disappear from (\ref{2}), leaving only a
dependence on Alice's measurement angles.

Remark: For clarity, in Fig.\ \ref{fig-1} we have assumed that the single
particle state associated with the first condensate has an orbital wave
function $u$ extending in space continuously from Alice's to Bob's
laboratory;similarly, the second condensate extends continuously from one
laboratory to the other.\ Nevertheless, we could also have assumed that any
(or both) of the wave functions is the coherent superposition of two
disconnected parts, one in region A and another in region B (as, for
instance, in Bell experiments where photons propagate to opposite detectors
without overlap of their wave packets).\ Provided both wave functions still
overlap in each measurement region, nothing is changed in the calculations
that will follow.\ The continuity of the condensates from region to the
other plays no role in the physical effect we discuss, illustrating that it
has nothing to do with some wave propagation along the condensates.

\subsection{Combined spin measurements}

The properties of formula (\ref{2}) have been discussed in \cite%
{Hidden-phase} and \cite{Mullin-Laloe}.\ All results of measurements have
exactly the same probability as for an ensemble of spins with well-defined
initial orientations defined by angle $\lambda $, with a random distribution
given by a function of this additional variable.\ Initially, the
\textquotedblleft additional\textquotedblright\ variable $\lambda $ has a
completely uniform distribution between $0$ and $2\pi $ and the first spin
measurement provides a completely random result.\ Then, while measurements
accumulate, the distribution of the \textquotedblleft
additional\textquotedblright\ variable becomes narrower and narrower. The
Bayes theorem can be used to show that the $\lambda $ distribution after $Q$
measurements is nothing by the product of cosines and sines squared that
appears under the integral of (\ref{2}); so this distribution is known at
any step of the experiment. At some point, this distribution becomes very
narrow; Alice's measurements have practically determined the relative phase
between the condensates (with some uncertainty); if she makes further
measurements in the direction in which the spins now point, she obtains
results that are practically certain.

\begin{figure}[h]
\centering \includegraphics[width=3in]{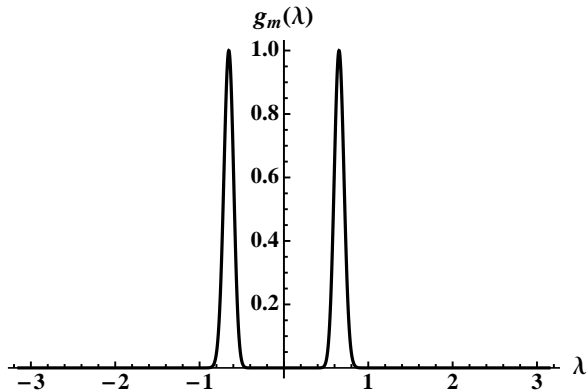}
\caption{A example of $g_{300}(\protect\lambda)$ Alice's probability
distribution after her first 300 measurements. }
\label{fig1}
\end{figure}

To illustrate the evolution of the $\lambda $ distribution, we have
performed simple numerical simulations using the same method as in \cite%
{Mullin-Krotkov-Laloe}. We rewrite Eq.\ (\ref{2}), the probability of
finding a $\pm$ spin along angle $\theta_{m}$ in the $m$th measurement, as
\begin{equation}
\mathcal{P}_{m}(\pm)\sim\int_{0}^{2\pi}\frac{d\lambda}{2\pi} g_{m}(\lambda)%
\left[1\pm\cos(\lambda-\theta_{m})\right]  \label{eq:ProbDist}
\end{equation}
In this equation, $g_{m}(\lambda)$ is the $\lambda$ distribution resulting
from the $m-1$ previous measurements:
\begin{equation}
g_{m}(\lambda)=\prod_{i=1}^{m-1}\left[1+\eta_{i}\cos(\lambda-\theta_{i})%
\right]
\end{equation}
where $\eta_{i}=\pm1$ is the result of the $i$th spin measurement result
made along angle $\theta_{i}$. The initial distribution ($m=1$) is perfectly
flat and the first result completely random. Then Alice starts a series of $%
300$ measurements along an arbitrary angle $\theta_{A}=0$, with random
results obtained from the $\lambda$ distribution at each step depending on
all previous results according to Eq.\ (\ref{eq:ProbDist}). In our
simulation she finds $n_{+}=269$ up spins and $n_{-}=31$ down spins. But as
the distribution of Fig. \ref{fig1} shows, at this stage she is unable to
tell whether the polarization that has developed during her measurements (or
that was already there) is at a positive or negative angle with respect to
her original zero angle. Note the sharpness of the distribution after this
relatively small number of measurements.
Alice then estimates that the polarization is at an angle
\begin{equation}
\lambda_{0}=2\cos^{-1}\left(\sqrt{\frac{269}{300}}\right)=\pm0.65
\end{equation}

\begin{figure}[h]
\centering \includegraphics[width=3in]{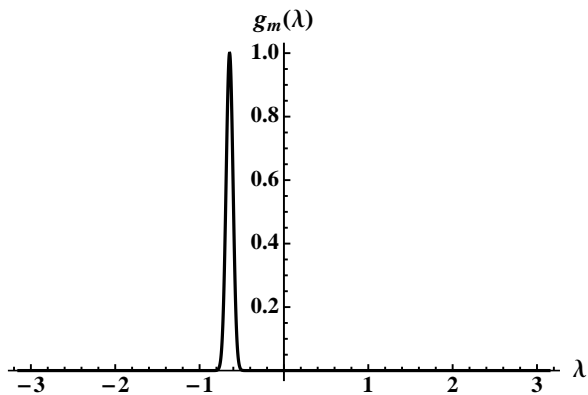}
\caption{A plot of $g_{600}(\protect\lambda)$, the total probability
distribution after Alice's second series of 300 measurements. She now knows
the polarization is near $\protect\lambda_{0}=-0.64.$}
\label{fig2}
\end{figure}

Then, to establish the sign of the polarization Alice measures at $%
\theta_{2}=\pi/2$ and finds $n_{+}=61$, $n_{-}=270$; this means that the
polarization angle is negative, with the estimate from this measurement
given by:
\begin{equation}
2\cos^{-1}\left(\sqrt{\frac{61}{300}}\right)=2.20
\end{equation}
relative to her new angle, or at $\lambda_{0}=\pi/2-2.43=-0.64$. The new
single-peaked distribution is shown in Fig.\ \ref{fig2}.

The same calculations apply to Bob's measurements. To make the argument easy
to follow we suppose that Bob picks his first set of angles also to be $%
\theta_{B}=0$; our simulation shows that he measures $n_{+}=269$, $n_{-}=31$%
, implying an angle of $\pm0.654$. If he does not know Alice's results, he
cannot determine its sign. He therefore makes a measurement at $\pi/2$ to
determine it, to find $n_{+}=58$, $n_{-}=242$ so that this estimate gives
the polarization angle as $2.23$ radians from his present angle, or at $%
\lambda_{0}\approx\pi/2-2.23=-0.66$. To check this result he measures 100
spins exactly at this angle and gets 99 spins up and only 1 down. The angle
found by Bob is very close to what Alice found.

\subsection{A non-local effect}

According to Eq. (\ref{2}), Alice and Bob share the same $\lambda $
distribution. This is the reason why, as we have just seen, if Bob starts
making his experiments when Alice has finished hers, he will necessarily
find a direction of $\lambda $ that falls inside the peak of the
distribution produced by Alice.\ But if he is not aware of the results
obtained by Alice, he will not notice anything special: to him, his results
appear to be the same as if no previous measurement had been made.\
Nevertheless, when both compare their results at the end of the experiment,
they can check that both have independently determined $\lambda $
distributions that are compatible: they then realize that Alice's results
have determined in advance the direction in which Bob has found the spins.\
In other words, Alice's measurements fix the relative phase of the two
condensates and therefore the direction of Bob's spins, which is doubly
paradoxical: first, if both laboratories are very remote, Alice cannot
interact with Bob's spins, which acquire a given direction by a non-local
effect; second, Alice may perform her experiments on a small number of
spins, $100$ or $1000$ for instance, while the spin orientation appearing in
Bob's laboratory may be macroscopic. How can a macroscopic physical quantity
appear under the effect of arbitrarily remote measurements performed on a
microscopic sample of atoms?

Another paradox arises when one asks the origin of the angular momentum
associated with the transverse orientation of Bob's spins, after Alice has
made her measurements (but before Bob started to make his). Since these
spins have interacted with no external physical system, one could expect
that their angular momentum should remain unchanged.\ How can a measurement
performed in Alice laboratory change the angular momentum contain in Bob's,
especially since this momentum is macroscopic? But quantum mechanics
predicts the values that Bob can observe when measuring this angular
momentum have indeed changed. The effect of Alice's measurement is
instantaneous, and has nothing to do with the propagation of any signal
along the condensates\footnote{%
As we remarked at the end of \S\ \ref{non-local-appearance}, the wave
functions do not necessarily extend continuously from one region to the
other; the phenomenon cannot be seen as due to some kind of vibration
propagating along the wave functions, even at infinite velocity.} (Bogolubov
phonons for instance); it is a pure non-local effect arising from the
quantum measurement postulate. The paradox is then that macroscopic angular
momentum seems to appear from nothing in Bob's laboratory.

\subsection{But does angular momentum exist before it is measured?}

\label{existence-angular-momentum}

Paraoanu and Healey \cite{Paraoanu-Healey}\ have studied the problem and
propose a way to solve the paradox.\ They state that, after Alice has made
her first measurements, "When Bob measures the spin-component on the large
cloud of condensate in his region, far from Alice, he is almost certain to
get a macroscopic result. But his does not mean that Alice's measurement
produces a `real' macroscopic value of angular momentum in Bob's region.\
Bob's measurement does not simply reveal this macroscopic value of a
pre-existing spin-component of or in the cloud produced by Alice's
projective measurement.\ Instead, the macroscopic spin-component `emerges'
during Bob's measurement following an interaction with Bob's measuring
device.... Certainly nothing that happens near Alice creates a macroscopic
angular momentum near Bob in violation of local conservation of angular
momentum\textquotedblright . The central point of their argument is
therefore that it is not possible to attribute physical reality to any
physical quantity (here angular momentum) prior to its measurement, even if
the state of the system is already an eigenstate of the corresponding
observable before measurement.

If no \textquotedblleft real\textquotedblright\ angular momentum exists in
Bob's laboratory before Bob measures it with his apparatus, and comes to
existence only during his measurement, it becomes natural to assume that
this apparatus takes the angular momentum recoil that is necessary to ensure
a local conservation of angular momentum. The purpose of the present article
is therefore to study the angular momentum recoil of both measurement
apparatuses, and examine if Bob's apparatus can take the recoil that is
necessary to ensure momentum conservation; if so, the paradox can be
lifted.\ Nevertheless, our analysis shows that this is not the case, so that
the paradox remains.

\section{Role of the measurement apparatuses}

\label{role-measurement-apparatus}

Before studying the paradoxical experiment, where the spin system is
initially in a double Fock state (double condensate), we discuss a simpler
case where the spin system starts from a phase state.\ In this case, a
macroscopic transverse spin orientation exists from the beginning of the
experiment.

\subsection{A preliminary experiment with a phase state}

\label{preliminary-expt}

Instead of (\ref{1}), we now assume that the initial state of the measured
system is a phase state, with the same total number of particles $%
N=N_{\alpha }+N_{\beta }$.\ This state depends on an arbitrary phase
parameter $\lambda _{0}$ and has the following expression:%
\begin{equation}
\left\vert \Psi \right\rangle =\frac{1}{\sqrt{N!}}\left[ e^{-i\lambda
_{0}/2}a_{1}^{\dagger }+e^{i\lambda _{0}/2}a_{2}^{\dagger }\right]
^{N}\left\vert 0\right\rangle  \label{3}
\end{equation}%
It can be seen as a single condensate where all $N$ particles are in the
same individual state:%
\begin{equation}
\left\vert \phi (\lambda _{0})\right\rangle =\frac{1}{\sqrt{2}}\left[
e^{-i\lambda _{0}/2}\left\vert u,+\right\rangle +e^{i\lambda
_{0}/2}\left\vert v,-\right\rangle \right]  \label{4}
\end{equation}%
where $\left\vert +\right\rangle $ and $\left\vert -\right\rangle $\ are the
single particle eigenstates of the $Oz$ component of the spin and $%
\left\vert u,v\right\rangle $ are the orbital states associated with the two
condensates.\ Actually these orbital states play no particular role in the
discussion \cite{Hidden-phase-Leiden}; we can assume that they coincide, or
even ignore them. The spin of the single particle state (\ref{4}) is fully
polarized in a transverse direction of the $xOy$ plane making azimuthal
angle $\lambda _{0}$ with axis $Ox$. In this case, the probability (\ref{2})
is replaced by:%
\begin{equation}
\begin{array}{l}
\displaystyle\mathcal{P}(m_{A}^{+},m_{A}^{-},m_{B}^{+},m_{B}^{-}|\varphi
_{A}^{1},..,\varphi _{A}^{P_{A}};\varphi _{B}^{k},..,\varphi _{B}^{P_{B}})=
\\
\multicolumn{1}{r}{\displaystyle=\prod_{j=1}^{m_{A}^{+}}\cos ^{2}\left(
\frac{\lambda _{0}-\varphi _{A}^{j}}{2}\right)
\prod_{j=m_{A}^{+}+1}^{m_{A}^{+}+m_{A}^{-}}\sin ^{2}\left( \frac{\lambda
_{0}-\varphi _{A}^{k}}{2}\right) \prod_{k=1}^{m_{B}^{+}}\cos ^{2}\left(
\frac{\lambda _{0}-\varphi _{B}^{k}}{2}\right)
\prod_{k=m_{B}^{+}+1}^{m_{B}^{+}+m_{B}^{-}}\sin ^{2}\left( \frac{\lambda
_{0}-\varphi _{B}^{k}}{2}\right)}%
\end{array}
\label{5}
\end{equation}%
The integral over $\lambda $ has now disappeared: all measurements\ are
independent, which is not surprising since all spins are initially polarized
in the same direction.

Consider a given spin measurement and the angular momentum transferred to
the measurement apparatus. Initially, the average angular momentum projected
along the measurement direction $\varphi _{A}^{j}$ is equal to $\left(
\hslash /2\right) \cos \left( \lambda _{0}-\varphi _{A}^{j}\right) $; after
measurement, it takes the value $+\hslash /2$ or $-\hslash /2$ randomly, but
on average it is:%
\begin{equation}
\frac{\hslash }{2}\left[ \cos ^{2}\left( \frac{\lambda _{0}-\varphi _{A}^{j}%
}{2}\right) -\sin ^{2}\left( \frac{\lambda _{0}-\varphi _{A}^{j}}{2}\right) %
\right] =\frac{\hslash }{2}\cos \left( \lambda _{0}-\varphi _{A}^{j}\right)
\label{6}
\end{equation}%
The projected angular momentum may be larger or smaller, depending on the
result of measurement, but on average it keeps the same value as before
measurement. The mean-square deviation $\Delta $\ of the angular momentum
transfer during $P$ identical measurements is given by:%
\begin{eqnarray}
\Delta ^{2} &=&P\left( \frac{\hslash }{2}\right) ^{2}\left\{ \cos ^{2}\left(
\frac{\lambda _{0}-\varphi _{A}^{j}}{2}\right) \left[ 1-\cos \left( \lambda
_{0}-\varphi _{A}^{j}\right) \right] ^{2}+\sin ^{2}\left( \frac{\lambda
_{0}-\varphi _{A}^{j}}{2}\right) \left[ -1-\cos \left( \lambda _{0}-\varphi
_{A}^{j}\right) \right] ^{2}\right\}  \notag \\
&=&P\left( \frac{\hslash }{2}\right) ^{2}\sin ^{2}\left( \lambda
_{0}-\varphi _{A}^{j}\right)  \label{7}
\end{eqnarray}%
As a consequence, on average over many spin measurements, the measurement
apparatuses takes zero angular momentum recoil.\ Nevertheless, this recoil
has a fluctuation that is proportional to the square root of the number of
realizations.

Without knowing the value of $\lambda _{0}$, Alice and Bob can determine it
through their measurements.\ In a first step, Alice can choose any
orientation $\varphi $\ for her measurements, and obtain an approximate
value of the cosine of the angle between this orientation and that of the
spins.\ She can then choose a perpendicular direction to obtain the sine of
this angle, removing the previous sign indeterminacy \cite%
{Mullin-Krotkov-Laloe}. After a few measurements, the best way to narrow the
distribution of possible values of $\lambda _{0}$ is to make spin
measurements in a direction perpendicular to her first estimation of $%
\lambda _{0}$. In this way, after a reasonable number of measurements, $100$
for instance, she has determined $\lambda _{0}$ with good accuracy.\
Finally, if she continues making spin measurements parallel to this
direction, almost no fluctuation remains in the results: these measurements
only confirm what she already knows, and perturb the spin system very
little.\ On the whole, Alice can determine the value of $\lambda _{0}$ by
transferring to her apparatus an angular momentum of the order of $10\hslash
$, maybe ten times more if she really needs an excellent accuracy, but she
certainly does not need to transfer a macroscopic number times of $\hslash $
to obtain the information.

Of course the same is true for Bob: to acquire information on the direction
of the (pre-existing) spin angular momentum in his laboratory, he may need
to transfer some angular momentum to his measurement apparatus, but again
not more than $10$ or $100$ times $\hslash $.

\subsection{Double condensate: angular momentum recoil of the apparatuses}

We now come back to the original thought experiment with the initial state (%
\ref{1}), where the initial angular momentum does not pre-exist
measurements, but is created by them. We must then use Eq.\ (\ref{2}) again,
and we assume that $N_{\alpha }=N_{\beta }$. As we have recalled above (\S\ %
\ref{non-local-appearance}), when Alice starts performing measurements, the $%
\lambda $ distribution is completely flat between $0$ and $2\pi $, and the
probabilities of the two results are equal (in contrast with \S\ \ref%
{preliminary-expt}).\ But, as soon as measurements become available, the $%
\lambda $ distribution becomes a product of sines and cosines, which becomes
a more and more peaked function.\ As above, we assume that Alice chooses appropriate
measurement angles to avoid sign ambiguities on $\lambda $ and to optimize
her definition of $\lambda $. After $100$ or $1000$ measurements, the $%
\lambda $ distribution has become a narrow peak around some value $\lambda
_{0}$, and one reaches a situation that is essentially the same as in \S\ %
\ref{preliminary-expt}.

Starting from this situation, Bob can obviously apply exactly the same
strategy as Alice; he does not know the value of $\lambda _{0}$, but can
measure it by transferring not more than $10$ or $100$ units of $\hslash $
to his measurement apparatus.\ After this preliminary operation, he can
choose a direction of measurement that is parallel to $\lambda _{0}$ and
make an arbitrary number of measurements, even macroscopic, without
transferring appreciable momentum to his apparatus. Even if he continues to
choose random directions of measurement, the angular momentum transfer
increases only as the square root of the number of measurements, not
linearly with it.\ So it is clear that the angular momentum transfer to his
apparatus remains totally negligible with respect to the measured spin
angular momentum.

We therefore see that, even if we consider that angular momentum does not
exist before it is measured (including the case where the state before
measurement is an eigenstate of the measured observable), the amount of
angular momentum appearing in Bob's laboratory cannot be compensated by an
opposite angular momentum transferred to his apparatus. Again, angular
momentum seems to have appeared from nowhere.

\subsection{Role of entanglement}

After Alice has performed at least one measurement, the two condensates
are strongly entangled.\ This is because her measurements relate to a coherent superposition of
the two spin states, so that they cannot distinguish
between spins of each condensate. The unmeasured spins are therefore left in a coherent
superposition of states where the population of each condensate fluctuates, with
their sum constant.

This raises the general question of quantum measurements performed on
entangled systems which, as we know, are not \textquotedblleft
separable\textquotedblright\ in quantum mechanics; they should be considered
as a whole.\ Could it be that, when a measurement is performed on a small
physical system that is entangled with another large system, the reaction of
the system on the measurement apparatus is completely changed by the
entanglement, becoming macroscopic instead of microscopic?

The epitome of an entangled states is a so called GHZ state
(\textquotedblleft all or nothing state\textquotedblright ), for which (\ref%
{1}) is replaced by:%
\begin{equation}
\left\vert \Psi \right\rangle _{0}=\frac{1}{\sqrt{N!}}\left[ \left(
a_{1}^{\dagger }\right) ^{N}+\left( a_{2}^{\dagger }\right) ^{N}\right]
\left\vert 0\right\rangle   \label{8}
\end{equation}%
With this state, if Alice measures the $Oz$ component of a single spin, all
subsequent measurements (by her or by Bob) will find the same result as the
first measurement: measuring the direction of one single spin
instantaneously fixes a parallel direction for all other spins.\
Nevertheless, before the first measurement, the average value of the $Oz$
total spin component vanished.\ Here again, angular momentum conservation
during the first measurement would imply that the measuring apparatus must take
the corresponding recoil. So one could conclude that, because Alice's spins
are part of a quantum system that is an indivisible whole, Alice's apparatus
takes a macroscopic angular momentum. We discuss this question further in
the next section.

\section{Discussion and conclusion}

Our conclusion is therefore that, even if we consider only situations where
angular momentum has been measured by Alice and Bob, the origin of the large
angular momentum that appears in Bob's laboratory cannot be found in his
measurement apparatus, which actually absorbs very little angular momentum.\
If we wish to preserve local conservation of angular momentum during
measurement processes, what are the possibilities to solve the paradox?

(i) A first possibility is to consider that angular momentum is not
necessarily conserved in each realization of the experiment, but only on
average.\ Indeed, the standard mathematical proof of angular momentum
conservation when $\mathbf{J}$ and $H$ commute shows that the average $%
\left\langle \mathbf{J}\right\rangle $ remains unchanged; the proof applies
to any power of $\mathbf{J}$, in other words that the complete distribution
of probability is constant.\ But the proof says nothing of individual events.

In this view, in quantum mechanics rotation invariance would not imply
angular momentum conservation for single experiments, but only statistically
for many realizations.

(ii) According to our analysis, it is not Bob's apparatus that takes the
angular momentum recoil, so could it be Alice's that takes the macroscopic
recoil?\ After all, it is her actions that apply state vector reduction and
transform the initial state vector (\ref{1}), with no transverse average
angular momentum, into (\ref{4}) where all spins are transversely polarized;
in other words, it is her measurements that \textquotedblleft
fuse\textquotedblright\ the two condensates into one and create a single
condensate described by a phase state with large angular momentum.\ In this
case, one should consider that, even if he directly interacts with $100$
spins only, since these spins are parts of (entangled with) a much larger
quantum system, her measurement relates to the whole system.

The problem with such a view is that it seems to imply instantaneous
signalling, creating a blatant contradiction with relativity.\ This is
because Alice's measurement could reveal if, for instance, Bob has destroyed
his condensate, or rotated its spin direction by applying a magnetic field. The same reasoning applies to the GHZ state discussed above. So this explanation does not seem appropriate.

(iii) On could argue that some \textquotedblleft super-selection
rule\textquotedblright\ forbids the preparation of such double Fock initial states.\
Clearly, gaseous spin condensates extending over long distances are
extremely fragile objects, even if repulsive interactions between the atoms
tend to stabilize them. One can therefore question the accessibility of such
double spin condensates.\ Nevertheless Bose-Einstein condensation seems to
provide a mechanism to generate such double Fock states in dilute gases,
since repulsive interactions favor the population of a single quantum state (%
\cite{Nozieres}), and this explanation sounds rather artificial.

(iv) Another explanation is to introduce additional variables, as suggested
in \cite{Hidden-phase}.\ In this case, quantum mechanics would not be
considered complete.\ In this perspective, transverse angular momentum would
exist from the beginning, and no paradox at all would occur. Nevertheless,
as discussed in \cite{Mullin-Laloe}, these additional variables should have
a component having non-local evolution since they can give rise to Bell
inequality violations.

In conclusion, the thought experiment we have discussed is not so remote
from experimentally accessible situations; the spontaneous appearance of
transverse spin polarization has already already been detected
experimentally \cite{Hall-et-al}.\ From a theoretical point of view, the
compatibility between quantum mechanics and the no-signalling constraint of
relativity has been the subject of many articles (for reviews, for instance
see \cite{Shimony-Stanford} or \cite{Seevinck}).\ The usual conclusion \cite%
{Shimony} is that quantum mechanics remains compatible with relativity, even
if parts of its formalism contains ingredients that are non-local and escape
space-time description (\textquotedblleft No story in spacetime can tell us
how nonlocal correlations happen\textquotedblright \cite{Gisin}).\
Nevertheless, the present work focuses, not on the formalism, but on the
results of measurements and their effects on measurement apparatuses, that
is macroscopic events that can presumably be considered as space-time
events.\ This makes the \textquotedblleft tension\textquotedblright with
relativity even more visible. It may be interesting to follow the line
opened by Wigner \cite{Wigner-1952} and examine in detail the general
question of how quantum systems react on measurement apparatuses, in
particular when they are entangled with another quantum system. This would
be useful to better understand how quantum mechanics manages to remain fully
compatible with relativity, and the reasons why a \textquotedblleft peaceful
coexistence between special relativity and quantum
mechanics\textquotedblright\ \cite{Shimony} can be maintained. A conclusion
of the present work is that, except if one prefers theories with non-local
additional variables, the simplest way to preserve non-signalling is that
proposed in (i) above: consider that angular momentum conservation applies
only statistically, but not in each realization of an experiment.

\end{document}